\documentclass[12pt]{article}
\usepackage{units}
\usepackage{bbm}
\usepackage{epsfig}
\usepackage{array}
\usepackage{float}
\usepackage{dsfont}
\usepackage{amstext}
\usepackage{rotating}
\usepackage{a4}
\usepackage{a4wide}
\usepackage{cite}
\usepackage{amsmath} 

\def\ra{\rightarrow}
\def\be{\begin{equation}}
\def\ee{\end{equation}}
\def\gs{\mathrel{
   \rlap{\raise 0.511ex \hbox{$>$}}{\lower 0.511ex \hbox{$\sim$}}}}
\def\ls{\mathrel{
   \rlap{\raise 0.511ex \hbox{$<$}}{\lower 0.511ex \hbox{$\sim$}}}}

\newcommand{\matrixx}[1]{\begin{pmatrix} #1 \end{pmatrix}}
\newcommand{\dd}{\mathrm{d}}
\newcommand{\ba}{\begin{array}{c}}
\newcommand{\baz}{\begin{array}{cc}}
\newcommand{\barrr}{\begin{array}{rrr}}
\newcommand{\bad}{\begin{array}{ccc}}
\newcommand{\bav}{\begin{array}{cccc}}
\newcommand{\baf}{\begin{array}{ccccc}}
\newcommand{\bea}{\begin{equation} \begin{array}{c}}
\newcommand{\eea}{ \end{array} \end{equation}}
\newcommand{\ea}{\end{array}}
\newcommand{\D}{\displaystyle}

\newcommand{\dm}{\mbox{$\Delta m^2$}}
\newcommand{\adm}{\mbox{$\overline{\Delta m^2}$}}

\newcommand{\del}{\partial}

\newcommand{\gsim}{\raise0.3ex\hbox{$\;>$\kern-0.75em\raise-1.1ex\hbox{
   $\sim\;$}}} 
\newcommand{\lsim}{\raise0.3ex\hbox{$\;<$\kern-0.75em\raise-1.1ex\hbox{
   $\sim\;$}}}

\begin{document}

\title{\vspace{-1cm}
\hfill {}\\[0.4in]
\vskip 0.4cm
\bf \Large
Gauged $L_\mu - L_\tau$ and different 
Muon Neutrino and Anti-Neutrino Oscillations: MINOS and beyond\\}
\author{
Julian Heeck\thanks{email: 
\tt julian.heeck@mpi-hd.mpg.de}~\mbox{
},~~Werner Rodejohann\thanks{email: 
\tt werner.rodejohann@mpi-hd.mpg.de}
\\\\
{\normalsize \it Max--Planck--Institut f\"ur Kernphysik,}\\
{\normalsize \it  Postfach 103980, D--69029 Heidelberg, Germany} 
}
\date{}
\maketitle
\thispagestyle{empty}
\vspace{0.8cm}
\begin{abstract}
\noindent  
If a $Z'$ gauge boson of a gauged $L_\mu - L_\tau$ symmetry is very
light, it is associated with a long-range leptonic force. In this case 
the particles in the Sun create via mixing of $Z'$ with the
Standard Model $Z$ a flavor-dependent potential for 
muon neutrinos in terrestrial long-baseline experiments. The 
potential changes sign
for anti-neutrinos and hence can lead to apparent differences in
neutrino and anti-neutrino oscillations without introducing CP or CPT 
violation. This can for instance 
explain the recently found 
discrepancy in the survival probabilities of muon neutrinos and 
anti-neutrinos in the MINOS experiment. 
We obtain the associated parameters of 
gauged $L_\mu - L_\tau$ required to explain this anomaly. 
The consequences for future long-baseline experiments and for the
anomalous magnetic moment of the muon are discussed. The main feature
of our explanation is that atmospheric neutrino mixing has to be
non-maximal. Neutrino masses tend to be quasi-degenerate.

\end{abstract}

\newpage
\section{\label{sec:intro}Introduction}

Additional gauged $U(1)$ symmetries are a feature of many theories
beyond the Standard Model (for a review, see 
e.g.~Ref.~\cite{PL_review}). A large amount of interesting
phenomenology arises in such scenarios, including 
LHC physics, lepton flavor violation, dark matter, etc. 
Here we focus on a particularly interesting class of models, namely 
anomaly free $U(1)$ symmetries under which the SM is 
invariant. It was observed 
long ago \cite{zero} that with the particle content of
the Standard Model one can gauge one of the lepton flavor combinations 
$L_e - L_\mu$, $L_e - L_\tau$ or $L_\mu - L_\tau$ without 
introducing anomalies. If the gauge bosons associated with this 
$U(1)$ symmetry are very light, then long-range 
forces are introduced. In case the extra $U(1)$
corresponds to $L_e - L_\mu$ or $L_e - L_\tau$, the electrons in
the Sun or the Earth generate a potential acting on the neutrinos in
terrestrial experiments \cite{u1lbl0,u1lbl0a,u1lbl}. 
The flavor dependence of $L_e - L_\mu$ or
$L_e - L_\tau$ induces modifications to the neutrino oscillations and
therefore the coupling of the $U(1)$ can be constrained. 
The lack of a significant amount of muons in the Sun or Earth lead 
to the fact that the oscillation phenomenology of gauged 
$L_\mu - L_\tau$ with very light $Z'$ was never studied, 
though this symmetry was analyzed
with different phenomenology in mind \cite{a_mu,dm,lmlt0,CR,lmlt}.

In the present letter we note that the unavoidable 
$Z$--$Z'$ mixing in models with gauged  
$U(1)$ symmetries allows to put limits on the parameters associated with 
$L_\mu - L_\tau$. The flavor dependent 
potential generated by the $Z'$ has different sign for neutrinos and
anti-neutrinos and can therefore lead to seemingly different neutrino
and anti-neutrino parameters. 
We apply this to the recently found discrepancy in
the survival probabilities of muon neutrinos and anti-neutrinos 
by the MINOS collaboration \cite{minos}. 
In this long-baseline experiment, the 
results for the oscillation parameters in the 
neutrino and anti-neutrino running lead to different values, 
namely\footnote{This result is henceforth referred to as ``MINOS anomaly''.}  
\bea \label{eq:minos}
\dm = \left(2.35^{+0.11}_{-0.08}\right) 
\times 10^{-3} \, {\rm eV}^2 ~,~~\sin^2 2
\theta > 0.91 \, , \\ 
\adm = \left(3.36^{+0.45}_{-0.40}\right)  \times 10^{-3}\, {\rm eV}^2~,~~
 \sin^2 2 \overline{\theta} = 0.86 \pm 0.11 \, , 
\eea
for neutrinos and anti-neutrinos, respectively \cite{minos}. 
We will use here the impact of a long-range force associated with the $Z'$ of
gauged $L_\mu - L_\tau$ to explain this anomaly.  We obtain the
parameters ($Z$--$Z'$ mixing and gauge coupling) of the $U(1)$ and 
discuss in addition consequences for future long-baseline neutrino
oscillation experiments
and the anomalous magnetic moment of the muon. An interesting feature 
of our proposal is that in order for gauged 
$L_\mu - L_\tau$ to be the explanation of the MINOS results,
atmospheric neutrino mixing needs to be non-maximal. We furthermore
find an interesting correlation in what regards the sign of the 
differences between neutrino and anti-neutrino 
parameters. Neutrino masses tend to be quasi-degenerate. 

Previous possible explanations for the MINOS results are CPT violation
\cite{cpt}, sterile neutrinos plus a gauged $B-L$ symmetry with a
massive ($\sim$ eV scale) $Z'$ \cite{sterile}, 
or non-standard interactions \cite{NSI}. 
The first two papers \cite{cpt} and \cite{sterile} were motivated by
previous low statistics results from MINOS, while Ref.~\cite{NSI} and
the present work use the recent higher statistics data sets \cite{minos}.\\ 

In Section \ref{sec:sce} we outline the framework of gauged 
$L_\mu - L_\tau$ symmetry including $Z$--$Z'$ mixing, current constraints are described in Section~\ref{sec:bounds}.
The results are applied to oscillation phenomenology and the MINOS
results in Section \ref{sec:minos}, where we also study the impact on 
future neutrino oscillation experiments, the anomalous magnetic
moment of the muon and neutrino masses. 
Section \ref{sec:concl} summarizes our findings.

\section{\label{sec:sce}Gauged $L_\mu - L_\tau$ Symmetry}
The most general Lagrangian after breaking the 
$SU(3)\times SU(2)\times U(1)_Y \times U(1)_{L_\mu - L_\tau}$ 
symmetry can be written as~\cite{ZZ'}
\begin{align}
{\cal L} = {\cal L}_\text{SM} + {\cal L}_{Z'} + {\cal L}_\text{mix}\,
, \label{eq:lagrangian}
\end{align}
where the relevant part of the Standard Model Lagrangian is
\begin{align}
	{\cal L}_\text{SM} = -\frac{1}{4} \hat{B}_{\mu\nu} \, 
\hat{B}^{\mu\nu} -\frac{1}{4} \hat{W}^a_{\mu\nu} \, \hat{W}^{a\mu\nu} +
\frac{1}{2} \hat{M}_Z^2 \, \hat{Z}_\mu \, \hat{Z}^\mu  -
\frac{\hat{e}}{\hat{c}_W} j_B^\mu \, \hat{B}_\mu
-\frac{\hat{e}}{\hat{s}_W} j_W^{a\mu} \, \hat{W}^a_\mu \, , 
\end{align}
and the hats denote that we are not in the mass eigenbasis. 
The currents $j_B^\mu$ and $j_W^{a\mu}$ are the usual Standard Model
ones. The gauge coupling of the 
$U(1)_{L_\mu - L_\tau}$ is denoted $\hat{g}'$.
The $Z'$ part in our case is
\begin{align}
	{\cal L}_{Z'} &= -\frac{1}{4} \hat{Z}'_{\mu\nu}\, 
\hat{Z}'^{\mu\nu}+ \frac{1}{2} \hat{M}_Z'^2 \, \hat{Z}'_\mu \, \hat{Z}'^\mu
- \hat{g}' \, j'^\mu \, \hat{Z}'_\mu \, , \\
	j'^\mu &= \bar{\mu} \, \gamma^\mu \, \mu 
+ \bar{\nu}_\mu \, \gamma^\mu
\, P_L \, \nu_\mu - \bar{\tau} \, \gamma^\mu \, \tau 
- \bar{\nu}_\tau \, \gamma^\mu\, P_L \, \nu_\tau \, , 
\end{align}
with the projection operator $P_L \equiv \frac{1}{2}(1 - \gamma_5)$. 
The term $\frac{1}{2} \hat{M}_Z'^2 \, \hat{Z}'_\mu \, \hat{Z}'^\mu$
breaks the $U(1)_{L_\mu - L_\tau}$ symmetry, and is generated by a vev of some 
Higgs sector (left unspecified here). 
Then there are terms associated with mixing of the field strength
tensors and the two massive bosons: 
\begin{align}  \label{eq:Lmix}
{\cal L}_\text{mix} = -\frac{\sin \chi}{2} \, \hat{Z}'^{\mu\nu}\, 
\hat{B}_{\mu\nu} + \delta \hat{M}^2 \, \hat{Z}'_\mu \, \hat{Z}^\mu
\end{align}
with the kinetic mixing angle $\chi$. The crucial mixing term
$\sin \chi$ can arise directly, or can be generated radiatively
\cite{holdom}. 

Diagonalizing \cite{ZZ'} the kinetic terms (which gives fields 
denoted by $B^\mu = \hat{B}^\mu + \sin \chi \, \hat{Z}'_\mu$ and
$Z'_\mu = \cos \chi \, \hat{Z}'_\mu$) 
and then the mass terms leads, besides the usual $W$ bosons, to a massless
photon field $A^\mu = \hat c_W \, B^\mu + \hat s_W \, {W^3}^\mu$ 
and two massive gauge bosons $Z_1$ 
and $Z_2$.  
They are related to the original $\hat{Z}$  and $\hat{Z}'$ as 
\begin{align}
Z_1^\mu &= \cos \xi  
\left(  \hat{Z}^\mu - \hat{s}_W \, \sin \chi \,  \hat{Z}'_\mu
\right) + \sin \xi \, \cos \chi  \, \hat{Z}'_\mu \, , \\
Z_2^\mu & =   \cos \xi \, \cos \chi  \, \hat{Z}'_\mu - \sin \xi \left(
\hat{Z}^\mu - \hat{s}_W \, \sin \chi \,  \hat{Z}'_\mu
\right)
\, ,  
	\label{eq:massbasis}
\end{align}
where $\xi$ is a new mixing angle defined by
\begin{align} \label{eq:xi}
	\tan 2\xi = \frac{-2\cos\chi \, (\delta \hat{M}^2 + \hat{M}_Z^2\, 
\hat{s}_W \, \sin \chi)}{\hat{M}_{Z'}^2 - \hat{M}_Z^2 \, \cos^2 \chi +
\hat{M}_Z^2 \, \hat{s}_W^2 \, \sin^2 \chi + 2\delta \hat{M}^2 \,
\hat{s}_W \, \sin \chi} \, . 
\end{align}
The above physical particles $Z_1$ and $Z_2$ are in the literature
normally called $Z$ and $Z'$. We will follow this notation from now
on. Their masses are given by
\begin{align} \label{eq:M12}
	M_{1,2}^2 = \frac{a+c}{2} \pm \sqrt{b^2 + \left(\frac{a-c}{2}\right)^2}
\end{align}
with
\bea \label{eq:abc} \D 
	a = \hat{M}_Z^2 \, ,  
\qquad b = \hat{s}_W \, \tan \chi \, \hat{M}_Z^2 
+ \frac{\delta \hat{M}^2}{\cos \chi} \, , \\ \D
	c = \frac{1}{\cos^2 \chi} \left( \hat{M}_Z^2 \, \hat{s}^2_W
\, \sin^2 \chi + 2 \hat{s}_W \, \sin \chi \, \delta \hat{M}^2 +
\hat{M}_{Z'}^2\right)  .
\eea
The situation simplifies considerably if the $Z'$ is much lighter than
the $Z$, i.e., if $\chi \ll 1$ and $\delta \hat{M}^2 \ll \hat{M}_Z^2$  
are very small. In this case we have for the masses 
\begin{align}
M_1^2 \simeq \hat{M}_Z^2 ~,~~ M_2^2 \simeq c -\frac{b^2}{a-c}
\, , 
\end{align}
and the mixing angle is 
\begin{align}
	\xi \simeq 
\frac{1}{\cos\chi}\left(\hat{s}_W \, \sin \chi + \frac{\delta
\hat{M}^2}{\hat{M}_Z^2}\right) \simeq\hat{s}_W \, \chi + \frac{\delta
\hat{M}^2}{\hat{M}_Z^2} \, . 
	\label{eq:deltaM}
\end{align}
With this approximation the Lagrangians for the physical 
particles are\footnote{Here we 
defined the physical Weinberg angle as 
$s_W^2 \, c_W^2 = \frac{\pi \, \alpha (M_1)}{\sqrt{2} \, G_F \, M_1^2}$. 
This gives the identity $s_W \, c_W \, M_1 = \hat{s}_W \, \hat{c}_W \,
\hat{M}_Z$ and the neutral current coupling constant becomes 
$e/(\hat{s}_W \, \hat{c}_W) \simeq e/(s_W \, c_W) \, (1- \xi^2/2)$.}
\begin{align}
	{\cal L}_A &= -e \, (j_\text{EM})_\mu \, A^\mu \,, \nonumber\\
	{\cal L}_{Z_1} &=  - \left(\frac{ e}{{s}_W \, {c}_W} \left( (j_3)_\mu -
{s}_W^2 \, (j_\text{EM})_\mu \right)  + g' \, \xi \, (j')_\mu\right) 
Z_1^\mu \,,  \label{eq:L_Z_2} \\
	{\cal L}_{Z_2} &= -\left(g' \, (j')_\mu -  (\xi - s_W \, \chi)  \frac{
e}{{s}_W \, {c}_W} \left( (j_3)_\mu - {s}_W^2 \, (j_\text{EM})_\mu
\right)- e \, c_W \, \chi \, 
(j_\text{EM})_\mu\right) Z_2^\mu \, . \nonumber
\end{align}
The Lagrangian for the $A^\mu$ field is the canonical one and hence $\hat{e}
= e$. The other gauge coupling 
$g'$ is simply $\hat g'$.\\

\begin{figure}[bht]
\centering
\includegraphics[scale=0.8]{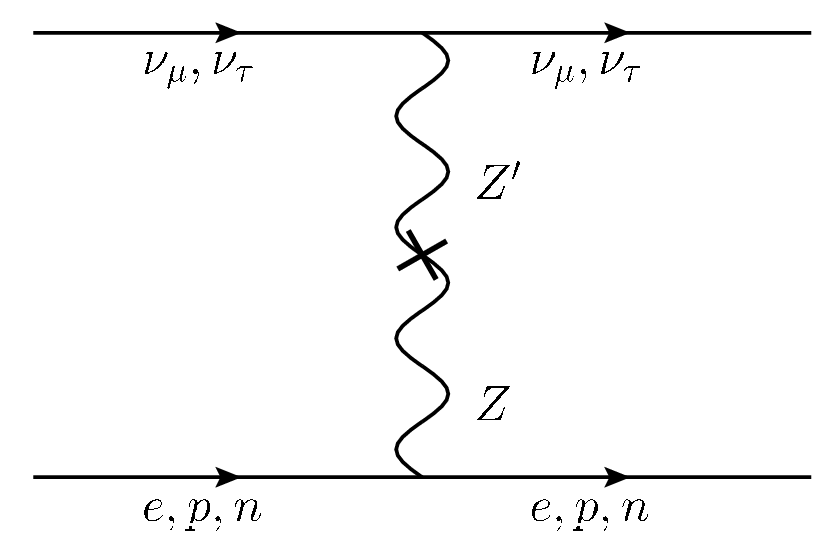}
\caption{Long-range $\nu_{\mu,\tau}$--$(e,p,n)$ interaction 
through $Z$--$Z'$-mixing.}
\label{fig:nu_scattering}
\end{figure}

If we take the mass of the $Z'$ to be $M_{2} < 1/R_{\rm A.U.} \simeq 
10^{-18}$ eV ($R_{\rm A.U.}\simeq 7.6\times 10^{26}$ GeV$^{-1}$~denotes 
an astronomical unit) we obtain for particles on Earth a static potential  
generated by particles in the Sun. 
This has been studied for the $U(1)_{L_e-L_\mu}$ and the 
$U(1)_{L_e -L_\tau}$ gauge bosons, for which the electrons in the Sun
generate a potential 
\be \label{eq:Vem} 
V = \alpha_{e\beta} \, \frac{N_e}{R_{\rm A.U.}} \simeq 
1.3 \cdot 10^{-11} \, \left(\frac{\alpha_{e\beta}}{10^{-50}} \right) {\rm eV} 
\ee
for the neutrinos $\stackrel{(-)}{\nu_\beta}$ on Earth. 
Here $\alpha_{e\beta} = g'^2/(4\pi)$ 
is the ``fine-structure constant'' of the $U(1)_{L_e -L_\beta}$ and 
$N_e$ is the number of electrons in the Sun. 
The constraints from solar 
neutrino and KamLAND data are $\alpha_{e\mu} 
< 3.4 \times 10^{-53}$ and $\alpha_{e\tau}<2.5\times 10^{-53}$ at
$3\sigma$~\cite{u1lbl0,u1lbl0a,u1lbl}. The lack of muons and taus
seems to forbid analogous studies of $L_\mu - L_\tau$, since its  
$Z'$ does not couple directly to protons, neutrons or
electrons. Consequently, to the best of our knowledge, 
there is no limit on $\alpha_{\mu\tau}$ from
oscillation experiments. 

However, there is an indirect effect due to the $Z$--$Z'$ mixing 
(see Fig.~\ref{fig:nu_scattering}). 
For a neutral and unpolarized Sun the final result for the potential
is (see the Appendix for details)
\begin{align} \label{eq:V}
	V_{\mu,\tau} = \pm\   g' \, (\xi -s_W \, \chi) \, 
\frac{e}{4 \, s_W \, c_W}
\frac{N_n}{4\pi R_{A.U.}} \, . 
\end{align}
Looking at Fig.~\ref{fig:nu_scattering}, the main features of 
this potential can be understood as $g'$ and $e/(s_W \, c_W)$ arising
from the vertices and $(\xi -s_W \, \chi)$ from the $Z$--$Z'$ mixing 
(see Eq.~(\ref{eq:L_Z_2})). 
The contributions of the electrons and protons cancel each other, so
that finally only the neutrons generate the potential. 
Their total number in the Sun is about $N_n \simeq N_e/4 \simeq 
1.5\times 10^{56}$. 
The Earth also generates a comparable potential, approximating a 
static potential at the surface, we get
\begin{align}
	\frac{V_\text{earth}}{V_\text{sun}} =
\frac{N_{n,\text{earth}}}{N_{n,\text{sun}}} \frac{R_{\rm
A.U.}}{R_\text{surface}} \simeq \frac{1.8\times 10^{51}}{1.5\times
10^{56}} \frac{1.5\times 10^8}{6380} \simeq 0.28 \, .
\end{align} 
Our full potential at the surface of the Earth is therefore:
\begin{align}
	V_{\mu,\tau} \equiv \pm V = \pm\  3.60\times 10^{-14} 
\, {\rm eV} \left(\frac{\alpha}{10^{-50}}\right) \mbox{ with } 
\alpha \equiv g' \, (\xi - s_W \, \chi) \, . 
	\label{eq:full_potential}
\end{align}
For anti-neutrinos, the sign of $V$ changes. We stress here that the
parameter $\alpha$ that we have defined is not a ``fine-structure
constant'' as for the $L_e - L_\mu$ or $L_e - L_\tau$ potentials, 
but a combination of coupling and mixing parameters. 
It can in particular be either positive or negative. 
Note further that due to the various factors in $V$ the scale for 
$\alpha = 10^{-50}$ is different than for 
$\alpha_{e\beta} = 10^{-50}$ in the cases of gauged $L_e -
L_\mu$ or $L_e - L_\tau$ in Eq.~(\ref{eq:Vem}).  
We will use in the following the value give in 
Eq.~(\ref{eq:full_potential}) for a
long-range force according to the Earth-Sun distance. In order not to completely
spoil the successful oscillation phenomenology, $V$ should not 
become too close to $\Delta m^2/E \simeq 2.9 \times 10^{-12} \, ({\rm
GeV}/E)$ eV, where we took for $\Delta m^2$ the mean of the two 
mass-squared differences from Eq.~(\ref{eq:minos}). 

The crucial $Z$--$Z'$ mixing, and consequently the potential
(\ref{eq:V}), can only be avoided if for the Lagrangian in Eq.~(\ref{eq:Lmix}) 
${\cal L}_{\rm mix} = 0$ holds, i.e., if both $\chi$ and 
$\delta\hat{M}^2$ vanish.  
As can be seen from 
Eq.~(\ref{eq:deltaM}), $\alpha$ would vanish for 
$\delta \hat{M}^2 =0$. In that case, however, one can show 
that the next order term 
for $\xi$ would generate non-zero $\alpha \simeq g' \, s_W \,
(M_{Z'}/M_Z)^2 \, \chi$, which is however too small for our purposes,
as we will see later. 
In the case $\chi = 0$,  the mixing angle is given by 
$\tan 2\xi = \frac{2 \, \delta \hat{M}^2}{\hat{M}_Z^2 -
\hat{M}_{Z'}^2}$, and $\alpha$ looks as before.

If $M_{2} < 1/R_\text{gal} \simeq 10^{-27}\, {\rm eV}$, with
$R_\text{gal}$ the distance between the Sun and the core of the 
galaxy ($R_\text{gal}\simeq 1.6\times 10^9 \, R_{\rm A.U.}$), 
we would obtain a potential 
\begin{align}
		\frac{V_\text{gal}}{V_\text{sun}} 
= \frac{(1-4)\times 10^{11}}{1.6\times 10^9}\simeq 60-240 \, ,
\end{align}
(with $100-400$ billion stars) which would dominate over the Earth and
Sun potentials. Depending on the range of the $U(1)$ force the results
which we obtain in the following can be easily rescaled. 

\section{\label{sec:bounds}Current bounds on $L_\mu-L_\tau$ parameters}

In this Section we will discuss the current bounds on the parameters
of $L_\mu - L_\tau$. They arise from gravitational fifth force
searches, electroweak precision observables, fermion charge
universality and cosmological considerations.   

In principle our model violates the equivalence principle because it
adds a lepton number dependent force to gravitation. The bounds on
such forces are very strict~\cite{EW} but are not directly applicable 
here since they are based on lunar ranging and torsion balance
experiments, which are only sensitive to the electron and baryon
content. The only effect comes once again from mixing; as shown 
in the Appendix, the potential corresponding to $Z'$ generated by 
a massive body depends on its neutron number $N_n$:
\begin{align}
	V (r)= \frac{e \, (\xi -s_W \, \chi)}{4 \, s_W \, c_W} \, 
N_n \, \frac{e^{-r  M_2}}{4\pi \, r} \, .
\end{align}
The gravitational potential between two bodies with masses $m_1$ and
$m_2$ and neutron content $N_{n_1}$ and $N_{n_2}$ is therefore changed to
\begin{align}
	V_\mathrm{grav} (r) = - G_N \frac{m_1 \, m_2}{r}
\left( 1 - \left( \frac{e \, (\xi -s_W \, \chi)}{4 \, s_W \, c_W}\right)^2 
\frac{N_{n_1}}{m_1}\frac{N_{n_2}}{m_2} \frac{1}{4\pi \, G_N} e^{-r  M_2}\right) .
	\label{eq:gravitational_potential}
\end{align}
The $95 \%$ C.L.~limits for a neutron dependent fifth force as a
function of its range are given in \cite{EW} (see references therein 
for a description of the experiments), where the effect of new light
vector or scalar bosons is parameterized as
\begin{align}
	V_\mathrm{grav} (r) = - G_N \frac{m_1 \, m_2}{r}\left( 1
+\tilde \alpha \frac{N_{n_1}}{\mu_1} 
\frac{N_{n_1}}{\mu_2} e^{-r/\lambda}\right) ,
\end{align}
$\mu$ being a test body mass in units of atomic mass unit $u$ and 
$\tilde \alpha = \pm \tilde g^2/(4\pi \, G_N \, u^2)$ (the sign 
distinguishes between vector and scalar interaction). 
Comparison with Eq.~\eqref{eq:gravitational_potential} gives the
translation into our parameters
\begin{align}
	|\tilde \alpha| \equiv \frac{1}{4\pi \, G_N \, u^2} 
\left( \frac{e \, (\xi -s_W \, \chi)}{4 \, s_W \, c_W}\right)^2 \, , &&
	\lambda \equiv \frac{1}{M_2} \, .
\end{align}
For Earth-Sun range we take the bound $|\tilde \alpha| < 10^{-11}$, 
given in~\cite{EW}, corresponding to 
\begin{align}
	(\xi -s_W \, \chi) < 5\times 10^{-24} \, ,
\end{align}
whereas the limit for an Earth range force is given as 
$|\tilde \alpha|<5\times 10^{-9}$, corresponding to
\begin{align}
	(\xi -s_W \, \chi) < 10^{-22} \, .
\end{align}
These are the strongest constraints on the mixing angles. 

The parameters are however also constrained 
through precision data from electroweak observables. 
Measurements around
the $Z$-pole examine the mass-eigenstate $Z_1$ with mass 
(see Eqs.~(\ref{eq:M12},\ref{eq:abc})) 
$M_1^2 \simeq a (1 + b^2/a^2)$, while 
measurements on $W$-bosons give values for $M_W = \hat M_Z \, 
c_W$. Therefore the mixing changes the $\rho$-parameter of 
the Standard Model from $\rho = M_W^2 / (M_Z^2 \, c_W^2)$ to
\begin{align}
        \rho_\mathrm{mix} = \left(\frac{M_W}{M_1 c_W}\right)^2 = \rho
\frac{1}{1+b^2/a^2} \simeq \rho \, (1-\xi^2) \, .
\end{align}
The current value~\cite{PDG} is $\rho = 0.9994\pm 0.0009$ which gives 
$\xi < 0.025$. Stronger limits arise by reading off from 
Eq.~(\ref{eq:L_Z_2}), the vector/axial couplings of the 
tauon: 
\begin{align}
	g_V^\tau \rightarrow 2 \, s_W^2 - \frac{1}{2} - 2 \frac{s_W \,
c_W}{e} g' \, \xi , && g_A^\tau \rightarrow -\frac{1}{2} \, ,
\end{align}
where $ 2 \, s_W^2 - \frac{1}{2}$ stems from the SM neutral current 
$ j_3^\mu - {s}_W^2 \, j_\text{EM}^\mu$. The asymmetry parameter 
$A^\tau \equiv 2 \, g^\tau_V \, g^\tau_A/((g^\tau_V)^2 + (g^\tau_A)^2)$ 
becomes approximately 
\begin{align}
	A^\tau \rightarrow A^\tau_{\rm SM} \left( 
1 + \frac{4 \, s_W \, c_W}{1 -
4 \, s_W^2} \frac{g' \, \xi}{e} \right) 
\equiv A^\tau_{\rm SM} + \Delta A^\tau (g' \,
\xi) \, , 
\end{align}
where $A^\tau_{\rm SM} = (1 - 4 \, s_W^2)/\left(1 - 4 \, s_W^2 
\, (1 - 2 \, s_W^2) \right)$ is the value without any new physics. 
This quantity is measured to be $A^\tau = 0.143\pm 0.004$
(Ref.~\cite{PDG}), while with the central value $\sin^2 \theta_W (M_Z)
= 0.23116$ one expects $A^\tau_{\rm SM} = 0.1499$. Since the measured 
$A^\tau$ and $A^\mu$ are of the same order while a nonzero $g' \, \xi$
shifts them in different directions, we will require  
$\Delta A^\tau (g' \, \xi)$ to be within the measured error, 
i.e.~$\Delta A^\tau (g' \, \xi) < 0.004$. This restricts $g'\,\xi$ 
to values
\begin{align}
	 g' \, \xi < 3.6 \times 10^{-4} \, .
\end{align}
This limit is stronger than e.g.~from the $Z$-coupling to $\nu_\mu$ or
the ratio $\Gamma (Z\ra \mu^+\mu^-)/\Gamma (Z\ra e^+ e^-)$, where
\begin{align}
	\Gamma (Z\ra \ell \bar \ell) = \frac{\alpha M_Z}{12 \, s_W^2
\, c_W^2} 
\left((g^\ell_V)^2 + (g^\ell_A)^2 \right)
\end{align}
at tree-level, ignoring lepton masses.\\

The mixing also changes the electromagnetic behavior, as can be seen 
from the Lagrangian~\eqref{eq:L_Z_2}, slightly rewritten and 
shown only for negatively charged muons ($\mu$), 
electrons ($e$) and positrons ($e^+$):
\begin{align} \nonumber 
	{\cal L}_{Z_2} = &-\left\{
	\left[g' + e \, c_W \, \chi -(\xi - s_W \, \chi) 
\frac{e}{s_W \, c_W} \left(s_W^2 - \frac{1}{4} \right) 
\right] \overline{\mu} \gamma_\beta \mu  \right. \\  
	& \left. - \left[e \, c_W \, \chi -(\xi - s_W \, \chi) 
\frac{e}{s_W \, c_W} 
\left(s_W^2-\frac{1}{4}\right)\right] \overline{e^+} \gamma_\beta e^+ 
\right. \\  
	& \left. + \left[e \, c_W \, \chi -(\xi - s_W \, \chi) 
\frac{e}{s_W \, c_W} 
\left(s_W^2-\frac{1}{4}\right)\right] \overline{e} \gamma_\beta e 
\right\} Z_2^\beta \nonumber 
\, , 
\end{align} 
In muonium the coupling between positive muons and electrons is 
modified because there is not only photon exchange, but also
photon-$Z'$ mixing. In direct analogy to the derivation of the
neutrino potential given in the Appendix, 
one finds an effective potential 
\begin{align}
V_{\mu^+ e^-} (r) = -\frac{e^2}{4\pi}\left(1 
- \frac{g'}{e} \, \tilde Q_P \, e^{-r M_2} \right)\frac{1}{r} \, .
\end{align}
where $\tilde Q_P \equiv -(\xi - s_W \, \chi) (1/4 -
s_W^2)/(s_W \, c_W) - c_W \chi$. 
Hence, the result is an effective change of the fine-structure 
constant in systems involving muons (or tauons)\footnote{There is an
effect quadratic in $\tilde Q_P$ due to two mixings in systems like
positronium or hydrogen, which is however way to small to be observable.}.
On atomic scales the factor $ e^{-r M_2}$ can be omitted. By
comparing the above potential with the potential for positronium 
we find the ratio of the $\mu^+$ and positron charge 
\begin{align}
	\frac{Q(\mu^+)}{Q(e^+)} 
= \frac{\frac{e^2}{4\pi} \left( 1 - 
\frac{g'}{e} \tilde{Q}_P\right)}{\frac{e^2}{4\pi}} 
\simeq 1 - \frac{g'}{e} \, \tilde{Q}_P \, . 
\end{align}
This ratio has been measured via the muonium
hyperfine-structure~\cite{finemuon} to be 
$1$ with an accuracy of $10^{-7}$, corresponding to a limit 
\begin{align}
	g' \left( 3 \, s_W \, \chi + (1- 4 \, s_W^2) \, \xi \right) 
< 5\times 10^{-8} \, .
\end{align}
Note that, as it should, there is no effect in case of $\chi = \zeta
= 0$, i.e., when there is no photon-$Z'$ mixing. In case di-muonium 
(a bound state of $\mu^-$ and $\mu^+$ \cite{brodsky}) would be
produced, one could test the $Z'$ even in the limit of no mixing. \\

Another effect the new light $Z'$ would have is a contribution to the
effective number of degrees of freedom, potentially threatening for
instance the success of Big Bang Nucleosynthesis (BBN). Recent BBN 
measurements as well as other cosmological probes are compatible with 
about one extra degree of freedom \cite{cosmo}. 
Let us demand that the $Z'$ does not contribute. This means for the
case of BBN that it should enter equilibrium after weak interactions 
freeze out ($T \simeq $ MeV), and requires to consider the process 
$Z' \, Z' \ra\nu_{\mu,\tau}\, \nu_{\mu, \tau} $, whose rate goes as 
$(g'^2/(4\pi))^2 \, T$. Comparing this to the Hubble rate $H \simeq T^2/M_{\rm
Pl}$ gives the requirement 
$g'^2/(4\pi) \ls 10^{-11}$ \cite{BBN}. A constraint of similar size
has been estimated from Supernova 1987a \cite{SN}. An
upper limit of $g'^2/(4\pi) \ls 10^{-18}$ can be obtained with the
process $\gamma \, \mu \ra \, Z' \, \mu$, going with $g'^2/(4\pi) \, 
\alpha\, T$, and demanding that $Z'$ is not in equilibrium at $T =
m_\mu$ \cite{BBN1}.  \\

As expected, the largest constraints stem from the equivalence
principle and BBN. However, the small values of the $L_\mu - L_\tau$ parameters
required in order to give observable effects in oscillation
experiments are compatible with these limits.

\section{\label{sec:minos}MINOS and Beyond}
The potential $V$ in Eq.~(\ref{eq:full_potential}) generated by 
$L_\mu - L_\tau$ is flavor dependent, acts on the
$\mu$--$\tau$ part of the system, and has a different sign for neutrinos and
anti-neutrinos. Consequently it is a good candidate for an explanation
of the MINOS results, which seemingly give different mixing parameters
in the muon neutrino and anti-neutrino oscillations. In a 2-flavor
approach, the Schr\"odinger-like equation for neutrinos is 
(note that we start in the mass basis) 
\begin{align} \label{eq:SE}
	i\frac{\dd}{\dd t} \vec{\nu}_M  = \frac{1}{2 E}\matrixx{m_2^2
&0\\ 0 & m_3^2} \vec{\nu}_M + V \, U^\dagger 
\matrixx{ 1 & 0\\ 0 & -1} U  \, \vec{\nu}_M \, , 
\end{align}
where $\Delta m^2 \equiv 
m_3^2 - m_2^2$ is the atmospheric mass-squared difference and 
$\vec{\nu}_M = (\nu_2, \nu_3)^T$ are the mass eigenstates which
are connected to the flavor states $\vec{\nu}_{\rm flavor} 
= (\nu_\mu, \nu_\tau)^T = U \, \vec{\nu}_M$ via the matrix 
\begin{align}
	U = \matrixx{\cos\theta & \sin\theta\\-\sin\theta &
\cos\theta} . 
\end{align}
Here $\theta = \theta_{23}$ is the atmospheric 
mixing angle.  

\begin{figure}[th]
\centering
\includegraphics[width=8cm,height=6cm]{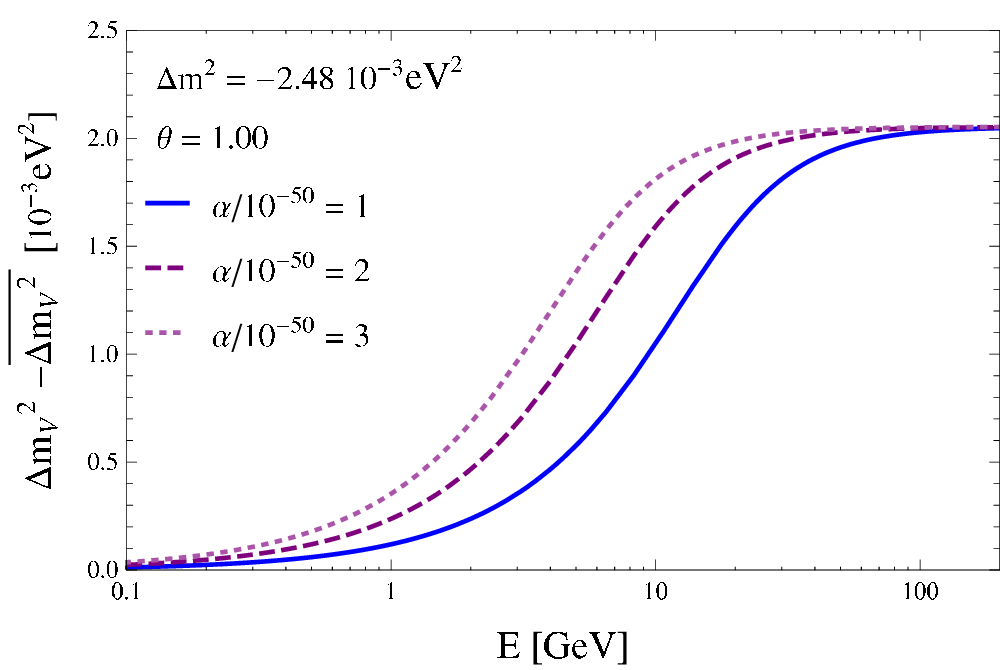}
\caption{Difference between the mass-squared
differences of neutrinos and anti-neutrinos (choosing initial values 
of $\dm = -2.48 \times 10^{-3}$ eV$^2$ and $\theta = 1$) for different values of 
$\alpha$ as a function of energy.}
\label{fig:diffdm}
\end{figure}

The Schr\"odinger-like equation (\ref{eq:SE}) thus contains the 
Hamiltonian:  
\begin{align}
	H_V = \frac{1}{2E} \matrixx{m_2^2 + 2  \, E  \, V  \, \cos
2\theta & 2  \, E  \, V  \, \sin 2\theta\\ 2  \, E  \, V  \, \sin
2\theta & m_3^2 - 2  \, E  \, V  \, \cos 2 \theta} = \frac{1}{2 \, E}
\, U_V \matrixx{m_{2,V}^2 &0\\ 0 & m_{3,V}^2}U_V^\dagger \, . 
\end{align}
As we have indicated, $H_V$ is diagonalized by the rotation matrix
\begin{align}
	U_V = \matrixx{\cos\phi & \sin\phi\\-\sin\phi & \cos\phi},
~\mbox{ with } 
\tan 2\phi = \frac{2\, \eta \, \sin 2\theta}{1-2 \, \eta \, \cos 2\theta} \, .
	\label{eq:new_angle}
\end{align}
We have introduced $\eta \equiv \frac{2 \, E \, V}{\Delta m^2}$. 
The new mass eigenvalues $m_{2,V}^2$ and $m_{3,V}^2$ are 
associated to the new mass eigenstates 
$\vec{\nu}_{M,V} = (\nu_{2, V}, \nu_{3, V})^T$ via 
\begin{align}
	\vec{\nu}_{M,V} = U_V^\dagger \, \vec{\nu}_M = U_V^\dagger \, 
U^\dagger \ \vec{\nu}_\text{flavor} \, . 
\end{align}
Thus, in the presence of the potential $V$, the mixing angle between 
flavor and mass eigenstates becomes $\theta+\phi$ and $\Delta m^2$ changes to 
$\Delta m_V^2 \equiv m_{3,V}^2-m_{2,V}^2$.  The exact results for
the parameters are 
\begin{align} \label{eq:mix}
\sin^2 2 \theta_V & = \frac{\sin^2 2 \theta}{1 - 4 \, \eta \, \cos 2
\theta + 4 \, \eta^2} \, , \\  \label{eq:mass2}
 \Delta m_V^2 & = \Delta m^2 \sqrt{1 - 4 \, \eta \, \cos 2 \theta + 4 \,
\eta^2 } 
= \Delta m^2 \sqrt{\frac{\sin^2 2 \theta}{\sin^2 2 \theta_V}}\, . 
\end{align} 
For $V = 0$ the vacuum
results $\sin^2 2 \theta$ and $\Delta m^2$ are obtained. 
For anti-neutrinos, the potential $V$ and hence $\eta$ changes sign,
thereby an apparent difference between the oscillation parameters of 
neutrinos $(\Delta m^2_V, \theta)$ and anti-neutrinos 
$( \overline{\Delta m^2_V}, \overline{\theta})$ could arise. 
Fig.~\ref{fig:diffdm} shows the difference between the mass-squared
differences of neutrinos and anti-neutrinos (choosing an initial value
of $\dm = -2.48 \times 10^{-3}$ eV$^2$) for different values of 
$\alpha$ as a function of energy.

We note here three important properties following from 
Eqs.~(\ref{eq:mix}, \ref{eq:mass2}):
\begin{itemize}
\item first, the effect goes with
$\eta \, \cos 2 \theta$, and therefore it is absent if $\theta$ is
maximal. In this case the oscillation parameters $\theta$ and $\Delta
m^2$ would be the same for neutrinos and anti-neutrinos, but 
with a common offset compared to their values for $V = 0$. 
If the long-range force mediated by $L_\mu - L_\tau$ is
responsible for the MINOS anomaly, then the necessary $\theta \neq \pi/4$ is a 
possibility to disentangle it from any other proposed explanation
\cite{cpt,sterile,NSI}; 
\item  the second point is that the corrections to the
mixing angle and the mass-squared difference are correlated. For
positive $\Delta m^2$ and $\alpha$ the correction for 
$\sin^2 2 \theta$ goes in the
opposite direction as the correction of the $\Delta m^2$. 
Recalling that MINOS finds $\overline{\Delta m^2} >  \Delta m^2$ we
therefore predict for positive $\Delta m^2$ and $\alpha$ that 
$\sin^2 2 \theta > \sin^2 2 \overline{\theta}$, which is compatible
with the MINOS results (see Eq.~(\ref{eq:minos})), 
and can be checked with higher statistics data
sets. For negative $\Delta m^2$ and positive 
$\alpha$ the correction goes in the same
direction, and hence  $\sin^2 2 \theta < \sin^2 2 \overline{\theta}$; 
\item  
the third point is that the relative effect is expected to be slightly
larger for $\sin^ 2 2\theta$ than for $\Delta m^2$. 
\end{itemize}

We can estimate the magnitude of the parameter $\eta$ as 
\be
\eta \simeq 0.025 \left(\frac{\alpha}{10^{-50}} \right) 
\left(\frac{E}{\rm GeV} \right), 
\ee
which allows for not too high energies (note that at MINOS the oscillation dip
occurs at around $E \sim 1$ GeV) and for $\alpha$ around $10^{-50}$ 
(see the discussion after Eq.~(\ref{eq:full_potential})), $\eta$ is small and
can be used as an expansion parameter. 
As can be seen from~\eqref{eq:mix} and~\eqref{eq:mass2} 
the relative difference of the mass-squared differences is in this case
obtained as 
\be
\frac{\Delta m_V^2 - \overline{\Delta m_V^2}}{\Delta m^2}  \simeq 
- 4 \, \eta \, \cos 2 \theta \, ,
\ee
while for the mixing angle the result is:  
\be
\frac{\sin^2 2 \theta_V - \sin^2 2 \overline{\theta}_V}
{\sin^2 2 \theta}  \simeq 8 \, \eta \, 
\cos 2 \theta  \, . 
\ee
These expressions nicely confirm the three points mentioned above. 
The muon neutrino and anti-neutrino survival probabilities are  
\begin{align} \label{eq:osc_probn}
	P \equiv P (\nu_\mu \rightarrow \nu_\mu) 
& = 1 - \sin^2 2 \theta_V \, \sin^2 \frac{\Delta m_V^2}{4 \, E} L  \, , \\
	\overline{P} = P (\overline{\nu}_\mu \rightarrow \overline{\nu}_\mu) 
& = 	P (\nu_\mu \rightarrow \nu_\mu)(\alpha \leftrightarrow
-\alpha) \, , \label{eq:osc_proban}
\end{align}
which are subject to the following degeneracies 
\be \label{eq:degs}
P(\theta,\Delta m^2,\alpha) = P(\theta,-\Delta m^2,-\alpha) = 
P(\theta + \pi/2,\Delta m^2,-\alpha) = P(\theta + \pi/2,-\Delta
m^2,\alpha) \, . 
\ee
While the part discussed so far was rather general, we continue by
applying the formalism to the recently found MINOS results \cite{minos}. 
We have performed with the expressions (\ref{eq:osc_probn}, \ref{eq:osc_proban}) 
a $\chi^2$-fit to the MINOS data (given in bins of energy $E_i$) on
the ratio of observed events divided by the expectation for no
oscillations. 
This data was taken, as in
Ref.~\cite{NSI}, from the slides of the talk
referred to in our Ref.~\cite{minos}. In case of asymmetric errors, the
largest one was used and inserted in the $\chi^2$-function  
\be \label{eq:chi2}
\chi^2 (\theta,\Delta m^2, \alpha) = \sum\limits_i
\left( \frac{P(\theta, \Delta m^2, \alpha, E_i) - R_i}{\sigma_i^2} 
\right)^2 + \sum\limits_i 
\left( \frac{\overline{P}(\theta, \Delta m^2, \alpha, E_i) - \overline{R}_i}
{\overline{\sigma}_i^2} 
\right)^2 , 
\ee
where $P$ $(\overline{P})$ is the survival probability $P(\nu_\mu \rightarrow
\nu_\mu)$ from Eq.~(\ref{eq:osc_probn}) 
(from Eq.~(\ref{eq:osc_proban})), $R_i$ ($\overline{R}_i$) the
ratio of observed events relative to the no-oscillation expectation,
and $\sigma_i$ ($\overline{\sigma}_i$) the error for the neutrino 
(anti-neutrino) data set. 
\begin{figure}[t]
\centering
\includegraphics[width=8cm,height=6cm]{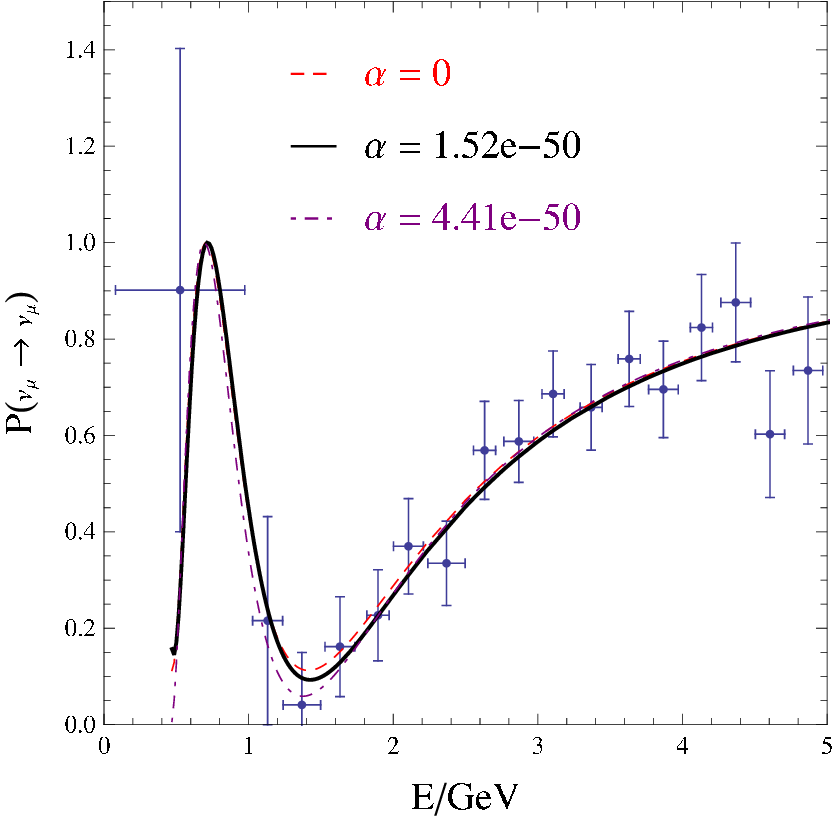}
\includegraphics[width=8cm,height=6cm]{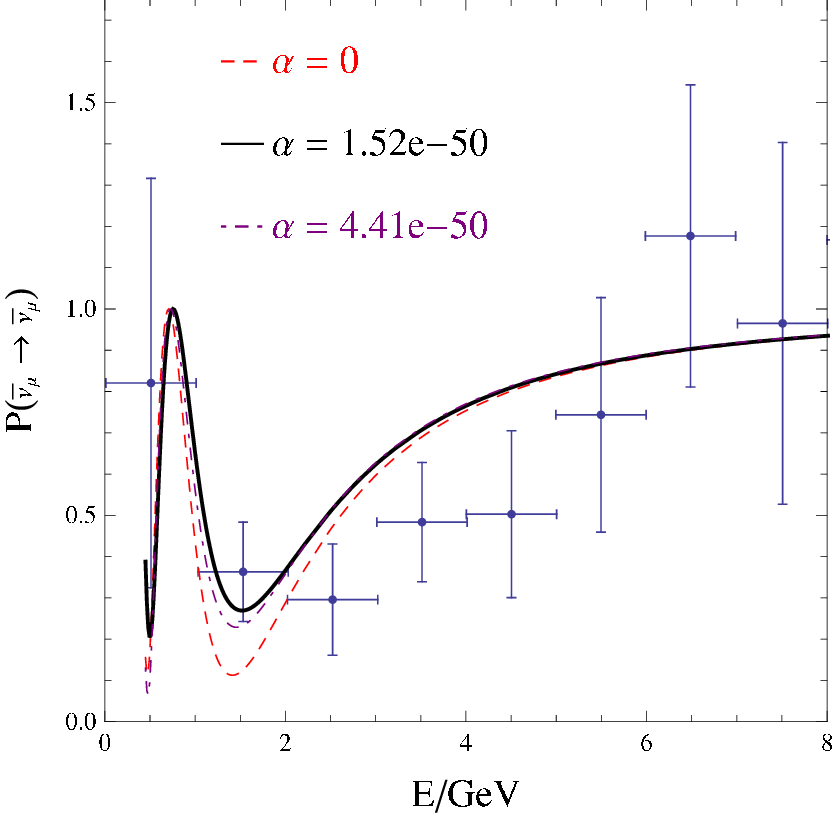}
\caption{The oscillation probabilities for the best-fit values from
Eq.~(\ref{eq:fit}) for neutrinos and
anti-neutrinos superimposed on the MINOS data. Also plotted are  
the cases $\alpha = 0$ and the value for the second, local 
$\chi^2$-minimum.}
\label{fig:probs}
\end{figure}
The result of our fit after marginalizing over $\dm$ and $\theta$ 
is\footnote{We have
checked our analysis by setting $\alpha = 0$ and have obtained the 
best-fit values $\dm = 2.28 \times 10^{-3}$ eV$^2$, 
$\sin^2 2 \theta = 0.94$ for the neutrino data set, and 
$\adm = 3.38 \times 10^{-3}$ eV$^2$, 
$\sin^2 2 {\overline \theta} = 0.81$ for the anti-neutrinos, in
good agreement with the MINOS results. A fit to the 
total data set yields $\dm = (2.38^{+0.20}_{-0.17}) \times 10^{-3}$ eV$^2$
and $\sin^2 2 \theta = 0.89^{+0.08}_{-0.07}$, with $\chi^2_{\rm min}/N_{\rm
dof} = 49.43/51 \simeq 0.97$.} 
\be \label{eq:fit}
\sin^2 2 \theta = 0.83 \pm 0.08~,~~
\Delta m^2 = (-2.48 \pm 0.19) \times 10^{-3}\,{\rm eV}^2~,~~
\alpha = \left(1.52_{-1.14}^{+1.17}\right) \times 10^{-50} \, , 
\ee
with $\chi^2_{\rm min}/N_{\rm dof} = 47.77/50 \simeq 0.96$. Recall the
degeneracies listed in Eq.~(\ref{eq:degs}). In Fig.~\ref{fig:probs} we
show the experimental data together with the results of our fit. 
\begin{figure}[t]
\centering
\includegraphics[width=5.26cm,height=5cm]{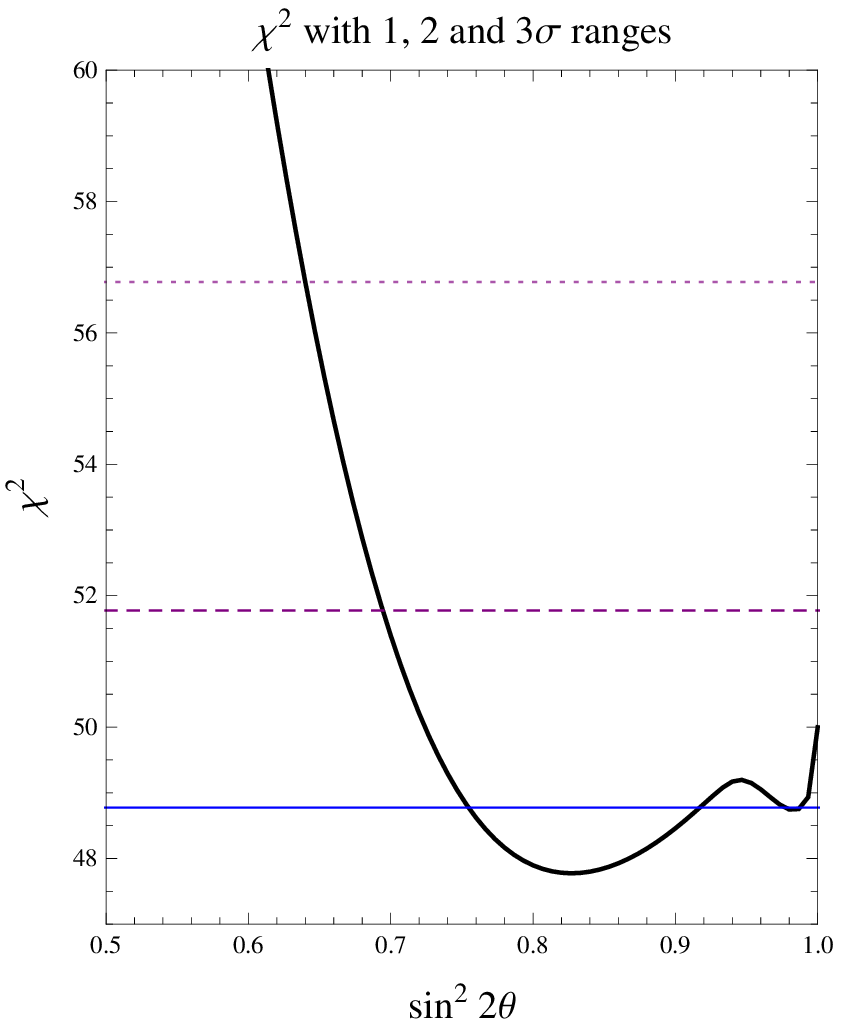}
\includegraphics[width=5.26cm,height=5cm]{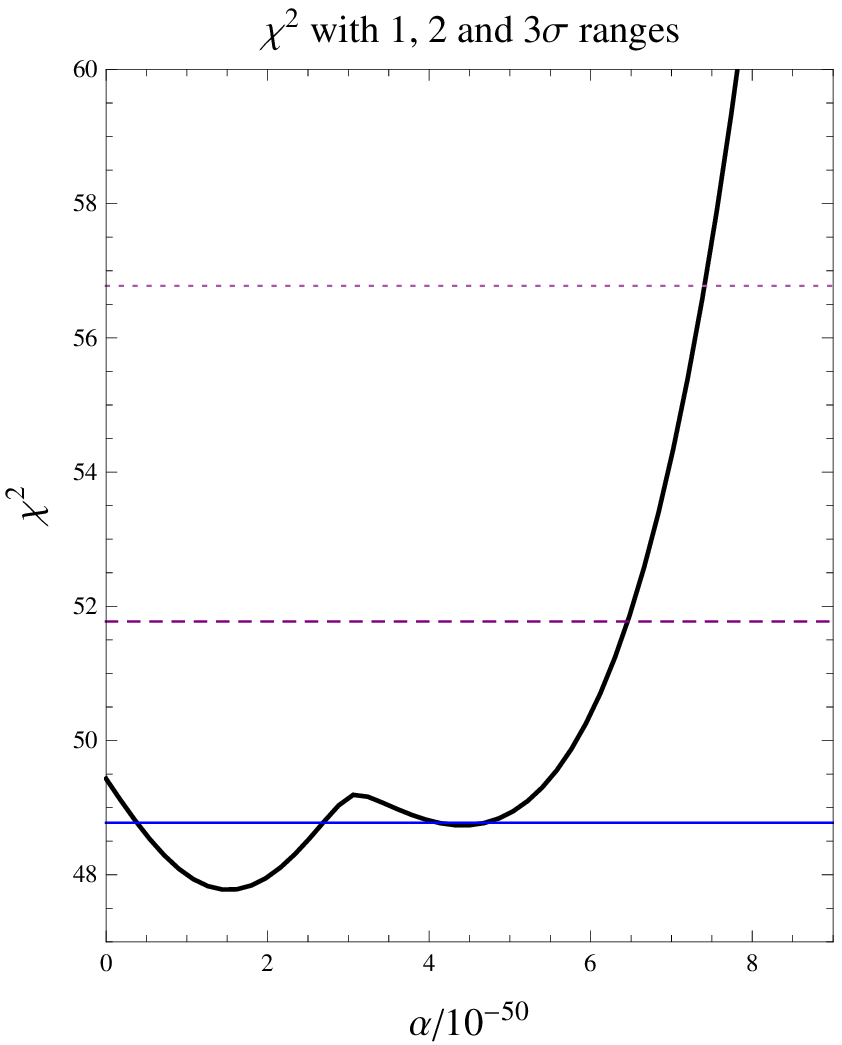}
\includegraphics[width=5.26cm,height=5cm]{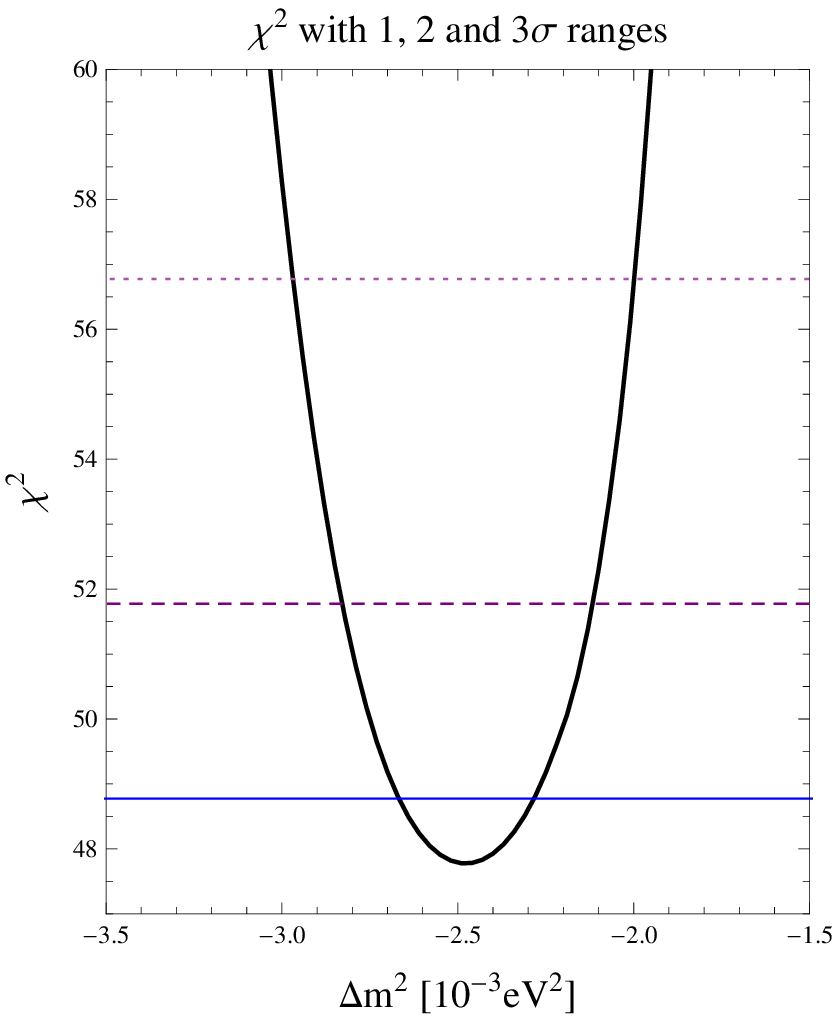}
\caption{The $\chi^2$-function from Eq.~(\ref{eq:chi2}) as a function
of the fit parameters.}
\label{fig:chi2}
\end{figure}
\begin{figure}[t]
\centering
\includegraphics[width=7cm,height=6cm]{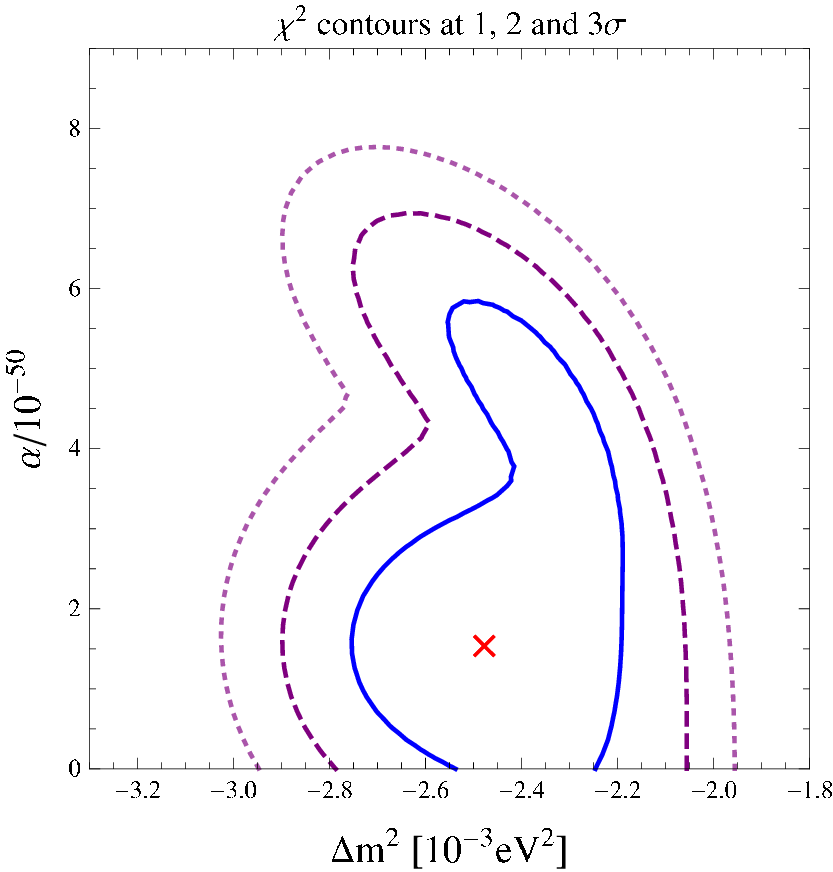}
\includegraphics[width=7cm,height=6cm]{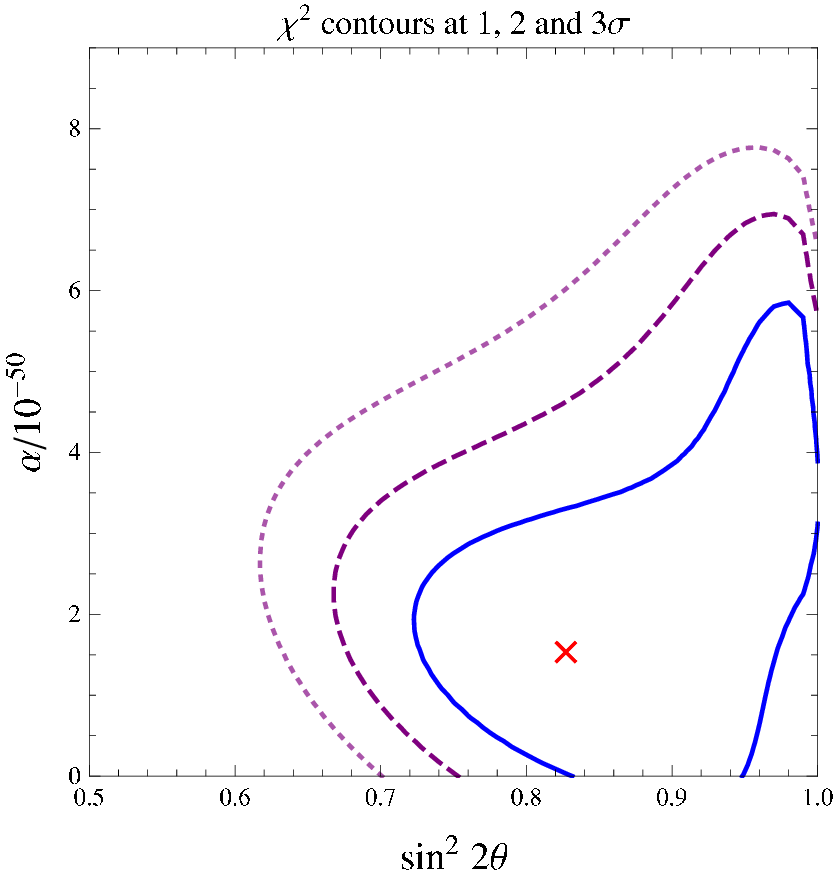}
\caption{The $1, 2$ and $3 \sigma$ 
(or equivalently $\Delta \chi^2 = 2.3,6.18, 11.83$) contours in the
$\alpha$--$\dm$ plane when marginalized over $\theta$ (left) and in
the $\alpha$--$\sin^2 2\theta$ plane when marginalized over $\dm$
(right). The cross marks the best-fit point.}
\label{fig:cont}
\end{figure}
One
can see that the non-zero value of $\alpha$ puts in particular 
the data points at the 
oscillation minimum in better agreement with the curves. From the 
plot of the $\chi^2$-function in Fig.~\ref{fig:chi2} one sees that there is a
second (local) minimum, corresponding to 
$\sin^2 2 \theta = 0.98$, 
$\Delta m^2 = 2.36 \times 10^{-3}\,{\rm eV}^2$ and 
$\alpha = 4.41 \times 10^{-50}$, with 
$\chi^2_{\rm min}/N_{\rm dof} = 48.73/50 \simeq 0.97$. The curves for
this point are also plotted in Fig.~\ref{fig:probs}. 
The second local minimum also explains the ``rabbit head looking'' 
shape of the contours in
$\alpha$--$\dm$ and $\alpha$--$\sin^ 2 2\theta$ space shown in
Fig.~\ref{fig:cont}.

The goodness of fit is not particularly worse
for the absence of new physics, which has been noted also 
in Ref.~\cite{NSI}. \\

\begin{figure}[t]
\centering
\includegraphics[width=5.26cm,height=5cm]{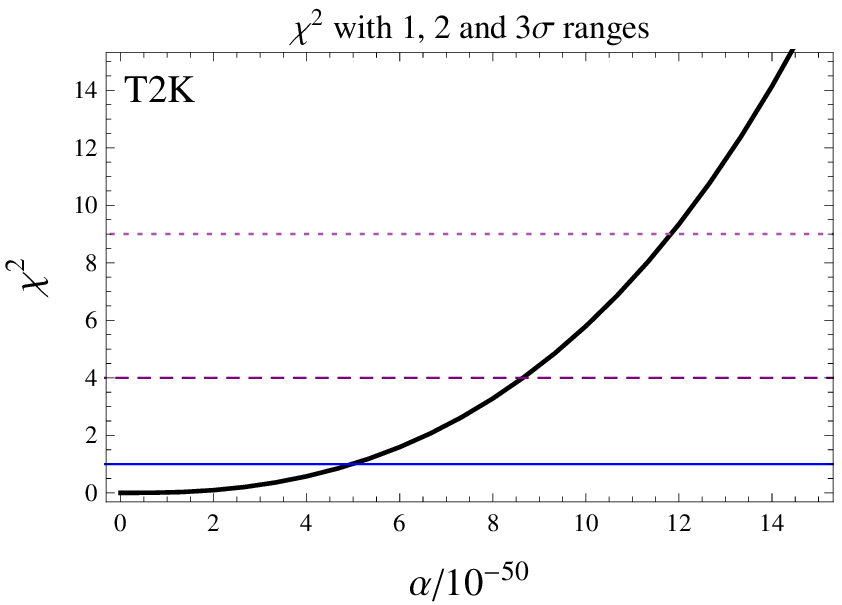}
\includegraphics[width=5.26cm,height=5cm]{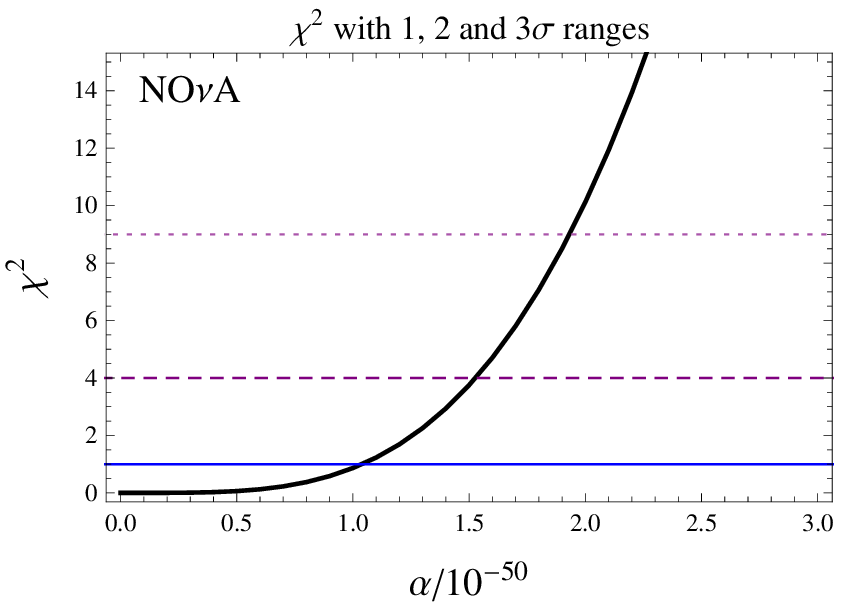}
\includegraphics[width=5.26cm,height=5cm]{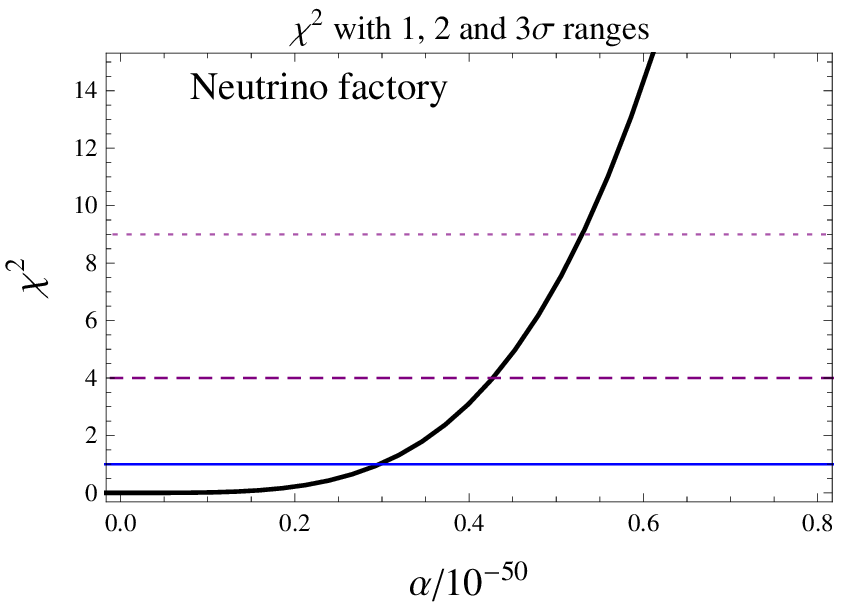}
\caption{The $1, 2$ and $3 \sigma$ limits which can be obtained by 
T2K (left), NO$\nu$A (middle) and a neutrino factory (right).}
\label{fig:globes}
\end{figure}

We continue by discussing the consequences of the implied value of
$\alpha$ in future neutrino oscillation experiments. We have 
modified the commonly used GLoBES software \cite{globes} to include
the potential $V$ from Eq.~(\ref{eq:V}). Using the pre-defined
packages (``AEDL files'') for the most frequently discussed 
future experiments, we analyzed T2K, NO$\nu$A and a neutrino factory, 
as listed in Table \ref{tab:LBLexps}, to obtain future constraints
on $\alpha$. The oscillation parameters we use are listed in Table
\ref{tab:osc}. The result is that at $3\sigma$, $\alpha$ can be constrained to be
below $11.80\times 10^{-50}$, $1.93\times 10^{-50}$ and $0.53\times
10^{-50}$, respectively. The $\chi^2$-functions generated by GLoBES
are shown in Fig.~\ref{fig:globes}.

\begin{figure}[t]
\centering
\includegraphics[width=5.26cm,height=5cm]{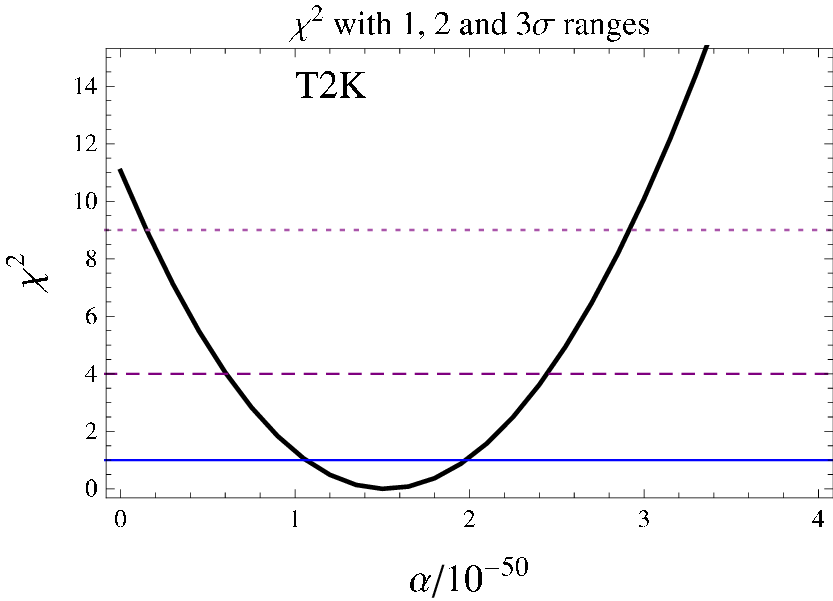}
\includegraphics[width=5.26cm,height=5cm]{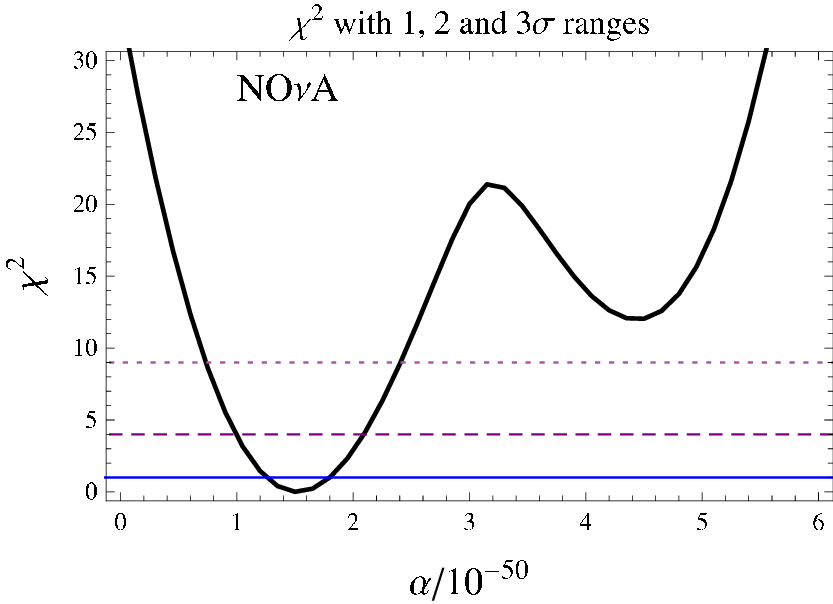}
\includegraphics[width=5.26cm,height=5cm]{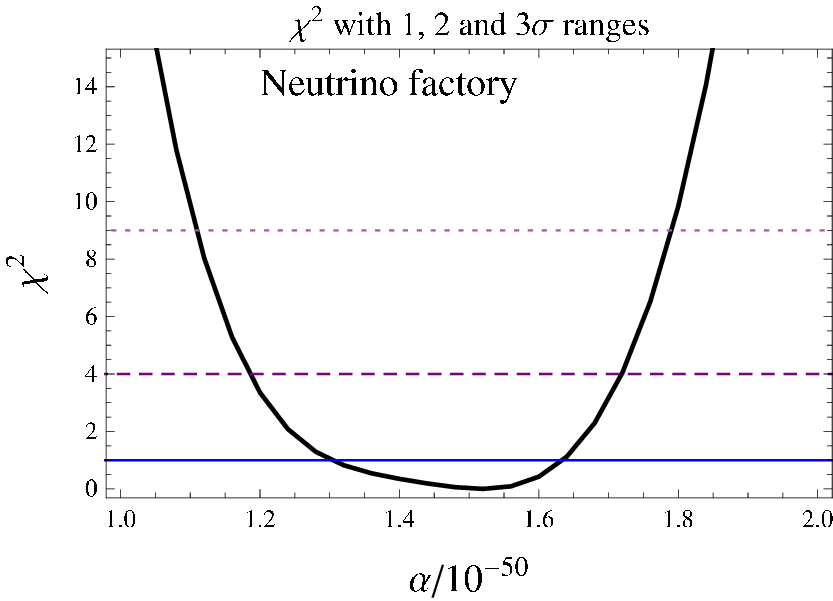}
\caption{The $1, 2$ and $3 \sigma$ constraints on 
$\alpha$ which can be obtained by 
T2K (left), NO$\nu$A (middle) and a neutrino factory (right) if
$\alpha = (1.52^{+0.11}_{-0.21})\times 10^{-50}$.}
\label{fig:globes2}
\end{figure}

Setting the true parameter values of $\alpha$, $\theta$ and $\Delta m^2$ (and their errors) to our
best-fit values from Eq.~(\ref{eq:fit}), we can see how the
``precision'' on $\alpha$ can be improved. From the 
plots of $\chi^2$ in Fig.~\ref{fig:globes2} one sees that NO$\nu$A would give 
$\alpha = (1.52\pm 0.27)\times 10^{-50}$, 
T2K would yield $\alpha = (1.52\pm 0.46)\times 10^{-50}$ and 
NuFact would determine very precisely  
$\alpha = (1.52^{+0.11}_{-0.21})\times 10^{-50}$.\\

As mentioned above, long-range forces generated by $L_e -
L_{\mu,\tau}$ have been discussed before. Ref.~\cite{u1lbl0} bounds 
$\alpha_{e\,\mu,\tau}$ by analyzing $\nu_\mu$ and $\nu_\tau$
oscillations and using atmospheric neutrino data. It is easy to see
that in a two-flavor framework, the potential $V_{e\tau} =
\alpha_{e\tau} \, N_e/R_{\rm A.U.}$ corresponds
to  $2 \, V_{\mu\tau}$. Likewise, $V_{e\mu}$ corresponds to 
$-2 \, V_{\mu\tau}$. Therefore, the 
limit of $\alpha_{e\tau} < 6.4\times 10^{-52}$ obtained in
Ref.~\cite{u1lbl0} corresponds to 
$\alpha = g' \, (\xi - s_W \, \chi) < 8.9 \times 10^{-50}$, not in
conflict with our fit-result from Eq.~(\ref{eq:fit}). In turn, this
means that not only $L_\mu - L_\tau$ could be the origin of the MINOS
anomaly, but also  $L_e - L_\mu$ or  $L_e - L_\tau$, for which 
$\alpha = 1.52 \times 10^{-50}$ translates into 
$\alpha_{e \, \mu,\tau} = 1.1 \times 10^{-52}$. 
If we take the $3\sigma$-bound $\alpha_{e\tau} < 2.5 \times 10^{-53}$
from solar neutrino and KamLAND data \cite{u1lbl0a}
and treat it like in the 2-flavor case we obtain 
$\alpha < 3.5 \times 10^{-51}$. However, the
interplay of the other limits on long-range forces, and also the
impact of stronger bounds on 
$\alpha_{e \, \mu,\tau}$ using solar and KamLAND data \cite{u1lbl0a}, 
can not be used without doing a full 3-flavor fit to all data. 
In general, we note that the different flavor structures of the
potentials arising from $L_e - L_\mu$, $L_e - L_\tau$ and $L_\mu -
L_\tau$,  
\be 
\left( 
\bad 
V & 0 & 0 \\ 
0 & -V & 0 \\
0 & 0 & 0 
\ea
\right)~,~~
\left( 
\bad 
V & 0 & 0 \\ 
0 & 0 & 0 \\
0 & 0 & -V 
\ea
\right)~,~~
\left( 
\bad 
0 & 0 & 0 \\ 
0 & V & 0 \\
0 & 0 & -V 
\ea
\right), 
\ee
render it difficult to translate existing bounds on  $L_e - L_\mu$ or 
$L_e - L_\tau$ into constraints on $L_\mu - L_\tau$, in particular if
in addition a matter potential is present in $V_{ee}$. We would like
to stress though that the solar neutrino oscillations should really be
fitted specifically for this model, since the electron and neutron densities in
the Sun are not proportional.

In Ref.~\cite{NSI} the presence of Non-Standard Interactions  
was assumed as the reason for the MINOS anomaly. In particular, a term 
$\epsilon_{\mu \tau}$ was introduced, and in the Hamiltonian it
appears together with the potential $V_{\rm m} \simeq \sqrt{2}  
G_F \, n_e \simeq 1.1 \times 10^{-13}$ eV. 
By fitting the MINOS data, the value 
$\epsilon_{\mu \tau} = -(0.12 \pm 0.21)$ was obtained. 
We note that for $|\epsilon_{\mu \tau}| = 0.1$ the term 
$V_{\rm m} \, \epsilon_{\mu \tau} $ is of
the same order of magnitude as our potential for $\alpha \simeq
10^{-50}$. A small difference to our explanation is that 
$V_{\rm m} \propto n_e$, i.e., the potential depends on the electron density,
which in turn depends on the matter density of the Earth. This changes
with baseline, and hence could in principle be used to distinguish 
Non-Standard Interactions from our explanation. We should note here that 
in a 2-neutrino framework the relation $2 \, V = 
V_{\rm m} \, \epsilon_{\mu \mu} $ holds, and our range of $\alpha$
would correspond to $\epsilon_{\mu \mu} \gs 0.25$, to be compared
with the 90 \% C.L.~limit \cite{nsi_lim} $|\epsilon_{\mu \mu}| \le
0.068$. Saturating this limit would correspond\footnote{Similar
comments apply for limits from atmospheric neutrino oscillations,
which have a somewhat stronger limit \cite{atm}.} to $\alpha = 1.04
\times 10^{-51}$. A fit to the data fixing it to this value yields 
$\sin^2 2 \theta = 0.88_{-0.07}^{+0.08}$ 
and $\dm = (-2.39^{+0.20}_{-0.17}) \times 10^{-3}$ eV$^2$, with 
$\chi^2_{\rm min}/N_{\rm dof} = 49.25/51 \simeq 0.97$. \\

It is worth discussing the anomalous magnetic moment of the
muon, where since many years a conflict between theory (i.e., its
Standard Model calculation) and experiment exists \cite{PDG}. 
The current experimental value of $a_\mu$ differs by $3.2\sigma$
from the Standard Model prediction, although there is some 
uncertainty in the hadronic contributions. 
Nevertheless, since the $Z'$ couples to the muon, it contributes to 
$\Delta a_\mu$ \cite{a_mu}.    
In the limit of $M_Z' \ll m_\mu$, the contribution is 
\begin{align}
	\Delta a_\mu &=\frac{g'^2 }{8\pi^2} \, , 
\end{align}
which in our light case translates into a constraint on the coupling
$g'$. From the constraint $\Delta a_\mu \ls 255\times 10^{-11}$ it
follows that $g' \ls 4.49\times 10^{-4}$. 
This would imply $(\xi - s_W \, \chi) \gs 3.3 \times 10^{-47}$ in
order to explain the MINOS anomaly.

Turning to neutrino masses, the conservation of $L_\mu - L_\tau$
dictates the effective 
neutrino Majorana mass matrix to be \cite{CR,lmlt}
\be \label{eq:mnu}
m_\nu = \left( \bad
a & 0 & 0 \\
\cdot & 0 & b \\
\cdot & \cdot & 0 
\ea \right) , 
\ee
regardless of its origin, such as some form of see-saw. 
It would result in neutrino masses $a$ and $\pm b$, hence 
one expects (close to) quasi-degenerate masses. Though the mass matrix
is $\mu$--$\tau$ symmetric, and hence implies $\theta_{13} = 0 $  and
$\theta_{23} = \pi/4$, it is too simple and can not reproduce all
data. Breaking $L_\mu - L_\tau$ is achieved by introducing extra Higgs
particles $\Phi'$, which obtain a vev. Necessarily, the implied scale
of the $Z'$ mass (which is generated by breaking of $L_\mu - L_\tau$) 
and the additional entries in $m_\nu$ are correlated via 
$m_Z' \sim g' \, \langle \Phi' \rangle$ and 
$(m_\nu)_{\alpha\beta} \ls \langle \Phi' \rangle$ if it is a weak
triplet, $(m_\nu)_{\alpha\beta} \ls v_{\rm wk} \,  
\langle \Phi' \rangle/\Lambda$ if it is a doublet and couples to the SM
Higgs, or $(m_\nu)_{\alpha\beta} \ls \langle \Phi' \rangle^2/\Lambda$
if it does not. Here $\Lambda$ denotes the high energy scale which
acts as the necessary suppression of the neutrino mass. 
Simultaneous ultra-light $Z'$ of order $10^{-19}$ eV and sizable 
$(m_\nu)_{\alpha\beta} \simeq 0.1$ eV implies for, 
say, $\Lambda = 10^{15}$ GeV that for doublets 
$\langle \Phi' \rangle$ is of order $10^3$ GeV and hence 
$g' \sim 10^{-30}$, while $g' \sim 10^{-17}$ for
triplets. Hence, the $Z'$ will essentially not contribute to the 
anomalous magnetic moment of the muon. 

\section{\label{sec:concl}Conclusions}
Long-range forces mediated by the $Z'$ boson associated with
gauged $L_\mu - L_\tau$ can lead to interesting and largely
unexplored phenomenology. For instance, 
neutrons in the Sun generate via $Z$--$Z'$ mixing a 
flavor-dependent potential for 
terrestrial muon and tau neutrinos. This potential changes sign for
anti-neutrinos, and hence can lead to apparent differences in neutrino
and anti-neutrino oscillations. Applying this new finding to 
the recently found MINOS anomaly implies a value of around $\alpha
\simeq 10^{-50}$, where $\alpha = g' \, (\xi - s_W \, \chi)$ is the
product of the new gauge coupling and the parameters quantifying the
$Z$--$Z'$ mixing. An interesting correlation between the
atmospheric neutrino parameters $\Delta m^2$ and $\theta$ is found. The latter
is required to be non-maximal, which is one of the handles to probe 
this explanation of the anomaly. 
By making use of the GLoBES software we have furthermore discussed 
future constraints on $\alpha$.  
Time will show whether the discrepancy in the MINOS results
survives. 
Nevertheless, many new physics effects imply 
different neutrino and anti-neutrino behavior, which underlines the
importance of analyzing them separately. The new effect arising from
$L_\mu - L_\tau$ (via $Z$--$Z'$ mixing) noted in the present letter is
one more example for this, and we have given simple estimates for
future constraints.

It would be interesting to discuss a similar approach for other
``anomalous'' oscillation results in which apparent differences of
neutrinos and anti-neutrinos are found, such as the recent 
MiniBooNE excess in a $\bar \nu_\mu \ra \bar \nu_e$ search
\cite{falsch}, or the slightly larger $\theta_{12}$ found in solar
neutrino analyses with respect to the $\theta_{12}$ in reactor
anti-neutrino experiments. As a final remark, neither CP nor CPT
violation are in this framework necessary for the apparent 
differences in neutrino and anti-neutrino oscillation probabilities.

\vspace{0.3cm}
\begin{center}
{\bf Acknowledgments}
\end{center} 
We thank Borut Bajc and Osamu Yasuda for helpful and interesting discussions. 
This work was supported by the ERC under the Starting Grant 
MANITOP and by the DFG in the project RO 2516/4-1 as well as in the 
Transregio 27.

\begin{table}[ht]
\centering
\begin{tabular}{|l|l|l|l|l|l|}
\hline
Experiment &  Baseline & Running-time [years]& Beam-energy $[\unit{GeV}]$ & Detector mass\\
\hline
\hline
T2K & $\unit[295]{km}$ & $5\ \nu$ $+5\ \overline{\nu}$ & $0.2-2$ & $\unit[22.5]{kt}$\\
NO$\nu$A & $\unit[812]{km}$ & $3\  \nu+3\ \overline{\nu}$ & $0.5-3.5$ & $\unit[15]{kt}$\\
Nufact & $\unit[3000]{km}$ & $4\  \nu+4\ \overline{\nu}$ & $4-50$ & $\unit[50]{kt}$\\
\hline
\end{tabular}
\caption{Parameters of long-baseline oscillation experiments simulated
by the GLoBES software \cite{globes}. \label{tab:LBLexps}}
\end{table}

\begin{table}[th]
\centering
\begin{tabular}{|l|l|}
\hline
$\theta_{12}$ & $\arcsin \sqrt{0.318} \pm 0.02\, (3\%)$\\
$\theta_{13}$ & $0 \pm 0.2$\\
$\theta_{23}$ & $\arcsin \sqrt{0.500} \pm 0.07\, (9\%)$\\
$\delta_\text{CP}$ & $\in [0,2\pi]$\\
$\Delta m_{21}^2\ [\unit[10^{-5}]{eV^2}]$ & $7.59\pm 0.23\, (3 \%)$\\
$\Delta m_{31}^2\ [\unit[10^{-3}]{eV^2}]$ & $2.40 \pm 0.12\, (5 \%)$\\
\hline
\end{tabular}
\caption{Oscillation parameters \cite{osc} 
used as input to the GLoBES simulation. 
\label{tab:osc}}
\end{table}

\renewcommand{\theequation}{A\arabic{equation}}
\setcounter{equation}{0}
\renewcommand{\thetable}{A\arabic{table}}
\setcounter{table}{0}

\begin{appendix}
\section{\label{sec:app}Derivation of the Potential}
For the sake of completeness, let us give here a derivation of the
static potential which the particles in the Sun generate for terrestrial
neutrinos. The potential (\ref{eq:Vem}) for gauged $L_e - L_\mu$ or 
$L_e - L_\tau$ can also be derived in this fashion. 
From Eq.~(\ref{eq:L_Z_2}) we consider the time-like 
components, note that $j_\text{EM}^0 = 0$ and have that  
\begin{align}
	j_3^0 = -\frac{1}{2} \, \bar{e}_L \, \gamma^0 \, e_L 
+ \frac{1}{2}  \, \bar{p}_L \, \gamma^0 \, p_L -\frac{1}{2}  \, \bar{n}_L \, 
\gamma^0 \, n_L  = -\frac{1}{4}\left(n_e - n_p +n_n\right) =
-\frac{n_n}{4} \, , 
\end{align}
since the axial-part will result in a spin-operator in the
non-relativistic limit and we assume the Sun is not polarized. 
The equation of motion for $Z_2^0$, following from the Euler-Lagrange equation
\begin{align}
	\del_\nu \frac{\delta}{\delta (\del_\nu  \, Z_{2
\mu})}\left(-\frac{1}{4}  \, {Z}_{2 \alpha\beta}  \,
{Z}_2^{\alpha\beta}\right) - \frac{\delta}{\delta  \, Z_{2
\mu}}\left(\frac{1}{2} {M}_{2}^2  \, {Z}_{2 \alpha}  \, {Z}_2^\alpha +
{\cal L}_{Z_2}\right) = 0 \, ,
\end{align}
is therefore
\begin{align}
	(\del^2 + M_{2}^2)  \, Z_2^0 = (\xi -s_W  \, \chi)
\frac{e}{s_W  \, c_W} \frac{n_n}{4}  \, .
\end{align}

In the static case outside of the Sun this is 
($n_n (\vec{x})= N_n  \, \delta^{(3)}(\vec{x})$):
\begin{align}
	(\Delta - M_{2}^2)  \, Z_2^0 = -(\xi -s_W  \, \chi) \frac{e}{s_W  \, c_W} \frac{1}{4}N_n  \, \delta^{(3)}(\vec{x})
\end{align}
with the well-known solution
\begin{align}
	V(r) = Z_2^0 = (\xi -s_W  \, \chi) \frac{e}{s_W  \,  c_W} \frac{1}{4}N_n \times \frac{e^{-r M_{2}}}{4\pi  \, r}  \, .
\end{align}
In the limit $M_{2}\rightarrow 0$ the potential, for $\nu_\mu$ and
$\nu_\tau$ respectively, on Earth is\footnote{We assume 
that the mixing angles are somewhat smaller than $g'$ so we can drop
the $\mathcal{O}(\xi^2,\chi^2,\xi \chi)$ terms against 
$\mathcal{O}(g' \chi, g'\xi)$. In the actual neutrino oscillation the
terms without $g'$ will be generation independent and therefore drop out.}:
\begin{align}
	V_{\mu,\tau} = \pm\   g' \, (\xi -s_W \, \chi) \frac{e}{4 \, s_W \, c_W}
\frac{N_n}{4\pi R_{\rm A.U.}} + \mathcal{O}(\xi^2,\chi^2,\xi \chi) \,
. 
\end{align}

\end{appendix}

\end{document}